\documentclass[prl,floatfix,twocolumn]{revtex4}
\usepackage{color,graphicx}
\usepackage{amsmath}
\usepackage{hyperref}

\begin{document}
\title[]{GPU implementation of mixed quantum-classical Liouville molecular dynamics without momentum jump
%    \footnote{
%        \href{https://doi.org/xx.xxxx/xxxx}{Chem. Phys. Lett. \textbf{xxx}, xxxxx (2026).}
%    }
}
\author{Koji Ando}
\affiliation{Department of Information and Sciences, 
    Tokyo Woman's Christian University, 
    2-6-1 Zenpukuji, Suginami-ku, Tokyo 167-8585, Japan}

\begin{abstract}
We implemented on GPU a mixed quantum-classical Liouville molecular dynamics simulation based on a momentum-jump-free theory.
The trajectory spawning that was previously implemented on CPU for sampling enhancement was eliminated
to avoid the overhead of thread divergence and dynamic memory allocation on the GPU.
The method was applied to a three-dimensional spin-boson model and a three-mode model of $S_2 \to S_1$ internal conversion in pyrazine involving a conical intersection.
A speedup of an order of magnitude compared to the CPU computation with spawning
was achieved,
as well as a linear scaling with respect to the number of sampling trajectories.
\end{abstract}

\maketitle

\section{ Introduction}
Nonadiabatic processes involving non-radiative transitions play a central role in fundamental chemical processes, including redox, acid-base, and photochemical reactions, across both gas and condensed phases, ranging from inorganic materials to biological macromolecules \cite{Nelson_non-adiabatic_2020,Li_iab_2021,Liu_nonadiabatic_2023}. Currently, the most sophisticated fully quantum-mechanical computational methods available include variational approaches such as the ML-MCTDH (multi-layer multi-configuration time-dependent Hartree) method \cite{Manthe2008_ML_MCTDH,Wang2015_ML_MCTDH_Review} and {non-Markovian and non-perturbative} frameworks such as the HEOM (hierarchical equations of motion) method for system-bath models \cite{Tanimura2020_HEOM,Lambert_qutip-bofin_2023}. Although these methods are expanding their applicability, each carries limitations in handling general multidimensional potential energy surfaces or bath spectral densities, as well as in covering wide range of temperature and system size.

In this context, semiclassical and mixed quantum-classical frameworks remain important. Among these, the surface hopping (SH) method \cite{Tully1990_FSSH,Wang_recent_2016,Jain_pedagogical_2022} has been the most practical and extensively adopted approach due to its comparative simplicity of implementation; nowadays, several software packages implementing variants of the SH method are readily available \cite{Mai_nonadiabatic_2018,Shakiba_libra_2022,Gardner_nqcdynamicsjl_2022}. Nevertheless, accurately capturing electronic decoherence remains a fundamental challenge within the SH paradigm, which has prompted the development of various decoherence correction schemes \cite{Granucci_including_2010,Tempelaar_generalization_2017,Plasser_strong_2019,Vindel-Zandbergen_study_2021,Shu_decoherence_2023}.

A naturally rigorous framework for incorporating electronic coherence relies on the quantum Liouville equation for the density matrix \cite{Donoso1998_JPCA,Kapral1999_MQCD,Santer2001,Ando2002_QCL,Ando2003_QCL,Shi_a_2004,Hanna_quantum-classical_2005,Wang_fewest_2015,Martens_surface_2019,Wu_a_2023} where off-diagonal elements of density matrix directly govern the coherence dynamics. In numerical simulations, however, capturing these dynamics within this formalism requires averaging over an immense ensemble of trajectories to resolve phase cancellations from the sign problem and achieve numerical convergence. Consequently, its widespread application has lagged behind the computationally more straightforward SH approach. Nonetheless, the density matrix formalism offers notable theoretical elegance and rigor, avoiding the reliance on ad hoc procedures.

To address this challenge, the present study implements a momentum-jump-free theory of mixed quantum-classical Liouville (QCL) molecular dynamics (MD) simulation \cite{Ando2002_QCL,Ando2003_QCL} on modern graphics processing unit (GPU) architectures in the Julia language \cite{Besard2019_TPDS}. 
This enables parallel computation over a total of $10^7$ to $10^8$ trajectories. 
By omitting trajectory spawning, a technique previously introduced in central processing unit (CPU) computations to enhance sampling, we reduce overheads related to thread divergence and dynamic memory allocation on GPUs, thereby achieving efficient massively parallel computation. 
Given the rapid rise of GPU acceleration in computational chemistry \cite{Seritan_terachem_2020,Kondratyuk_gpu-accelerated_2021,Manathunga_quantum_2023,Guo_byteqc_2025}, this work lays the groundwork for future integration with GPU-accelerated electronic structure algorithms and general MD simulations.

We also note that a promising variant of SH method, the mapping approach to surface hopping \cite{Richardson_nonadiabatic_2025}, 
has recently emerged, 
the accuracy of which appears to depend on reproducing the QCL time-correlation function to first order in time.

Section \ref{sec:method} outlines the basic theory and the GPU implementation. Section \ref{sec:results} presents and discusses the computational results. The final section concludes.

\section{ Theory and Computation}
\label{sec:method} 
\subsection{ Mixed QCL theory without momentum jump}
Here we briefly outline the theoretical framework established in Refs. \cite{Ando2002_QCL,Ando2003_QCL}. It was derived from first-principles under well-defined approximations, first taking the matrix element over the electronic basis and subsequently performing a partial Wigner transformation on the nuclear coordinates. Because the resulting equations of motion do not contain the terms representing off-diagonal Hellmann-Feynman forces, which were interpreted as momentum-jump operations during nonadiabatic transitions, the numerical simulation is free from energy divergence near classical turning points associated with the momentum-jump. As a result, improved numerical stability and comparable or better agreement with quantum reference calculations were obtained for one- and three-dimensional spin-boson models as well as a two-state three-mode model of $S_2 \rightarrow S_1$ internal conversion in pyrazine involving a conical intersection \cite{Ando2003_QCL}.

The equations of motion for the density matrix element $\rho_{\alpha \beta :\mathrm{W}}$ between 
{the adiabatic electronic} states $\alpha$ and $\beta$ after partial Wigner transformation on the nuclear degrees of freedom 
have been derived as \cite{Ando2002_QCL,Ando2003_QCL} 
\begin{multline}
    \frac{\partial}{\partial t} 
    \rho_{\alpha \beta :\text{W}}\left(R , P\right) = 
    - i \omega_{\alpha \beta}\left(R\right)\rho_{\alpha \beta :\mathrm{W}}\left(R , P\right) \\
    -\frac{P}{M} \sum_{\gamma} 
    \left(d_{\alpha \gamma}(R)\rho_{\gamma \beta :\mathrm{W}}(R , P)-\rho_{\alpha \gamma :\mathrm{W}}(R , P)d_{\gamma \beta}(R)\right) \\
    -\frac{P}{M} 
 \frac{\partial \rho_{\alpha \beta :\mathrm{W}}(R , P)}{\partial R} - \frac{1}{2} \left(F_{\alpha} (R)+ F_{\beta} (R)\right)
 \frac{\partial \rho_{\alpha \beta :\mathrm{W}}(R , P)}{\partial P}
\label{eq:eom}
\end{multline} 
in which $R$, $P$, and $M$ represent the coordinates, momenta, and masses of nuclear degrees of freedom, the subscripts $\alpha , \beta$, and $\gamma$ indicate the electronic states, $\hbar \omega_{\alpha \beta}$ is the energy difference between states $\alpha$ and $\beta$, $d_{\alpha \beta}$ is the first-order non-adiabatic coupling, and $F_{\alpha}$ is the classical force on the {adiabatic} potential energy surface of state $\alpha$. 
{Note that the equations do not include the terms representing the off-diagonal Hellmann-Feynman forces,
which comes from the operational order of the matrix element calculations and the partial Wigner transformation,
with the $R$-dependence of the adiabatic (or diabatic) electronic basis \cite{Ando2002_QCL,Ando2003_QCL}.}

The basic framework of the computational method is identical to that in Refs. \cite{Santer2001,Ando2003_QCL}, 
except that the present GPU implementation does not involve the trajectory spawning. 
{(See the next section \ref{sec:GPU_implementation}.)}
The CPU code with the spawning was rewritten in Julia.

\subsection{ GPU implementation without trajectory spawning}
\label{sec:GPU_implementation}
In the former CPU code, 
{new walkers are generated (spawned)} 
whenever the transition probability exceeded a random number, making the total number of walkers (trajectories) non-deterministic. By contrast, the number of walkers in the present GPU implementation is predetermined. State variables for all walkers, including density matrix elements, coordinates, and momenta, are pre-allocated in continuous GPU memory arrays. A GPU kernel is then launched at each time step to propagate the entire ensemble by one step.

{In short, the simulation proceeds as follows.
First, an ensemble of points (walkers) in phase-space $(R,P)$ is sampled randomly in the Wigner distribution at zero temperature.
To each walker, an index and a value of the density matrix element $\rho_{\alpha\beta}$ is assigned.
As the last two terms of Eq. (\ref{eq:eom}) indicate, 
the walkers with the index $\alpha\beta$ propagate classical mechanically with the force $(F_\alpha(R) + F_\beta(R))/2$.
At each simulation step, the transition probability is computed from the non-adiabatic coupling $d_{\alpha\beta}$.
If it exceeds a random number between 0 and 1, the walker is assigned another index;
for instance, from $\rho_{22}$ to either $\rho_{12}$ or $\rho_{21}$ with equal probabilities. 
Finally, the phase factor of $\exp(-i\omega_{\alpha\beta}t)$ is multiplied to the density matrix element,
before proceeding to the next time step.
}

The GPU code in Julia was implemented with the CUDA.jl library \cite{Besard2019_TPDS}. 
The number of threads per block was set to 256, and the grid size (number of blocks) was determined via ceiling division 
to ensure that a sufficient number of blocks was allocated to cover all the walkers. 
The reductions for ensemble average were performed with the \texttt{mapreduce} function 
rather than by the \texttt{@atomic} macro, as the former ensures numerical reproducibility 
by maintaining a deterministic order of summation. 
The CPU and GPU computations were in double and single precision, respectively. 
The calculations were performed on a workstation with Intel Xeon(R) Gold 5317 3.00 GHz CPU 
and {two} NVIDIA GeForce RTX 3090 GPU {cards}.

\subsection{ Simulation Models}
For numerical examination, we employed the same models as in Refs. \cite{Santer2001,Ando2003_QCL}, three-dimensional spin-boson model and a two-state three-mode model of the $S_2 \rightarrow S_1$ internal conversion of pyrazine. 
{(See, in particular, Sec. IV A and C of Ref.\cite{Santer2001} and Sec. 2 of Ref.\cite{Schneider_surface-hopping-induced_1989}.)}
We have also carried out calculations on the one-dimensional spin-boson model, 
but we omit the results {(except those in Fig. \ref{fig:fig5})} 
as the discussion and insights were consistent with those for the other two models.

\section{ Results and Discussion}
\label{sec:results} 

{The results in this section are from the single GPU computations,
except for the comparison with calculations with two GPUs noted at the end of Sec. \ref{sec:GPUscaling}.}

\subsection{ Accuracy and Convergence}

\begin{figure}[htbp]
\centering
 \includegraphics[width=0.45\textwidth]{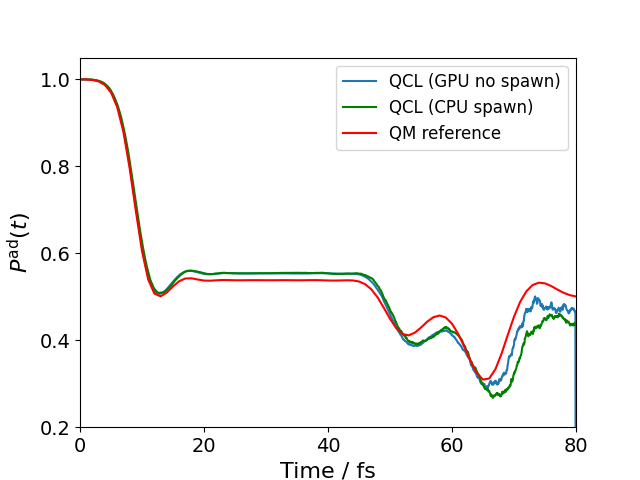}
 \caption{Adiabatic excited state population $P^{\text{ad}} \left(t\right)$ for three-mode spin-boson model, comparing results from GPU without trajectory spawning, CPU with trajectory spawning, and quantum reference calculation.}
 \label{fig:fig1}
\end{figure}

Figure 1 displays the adiabatic excited state population $P^{\text{ad}} \left(t\right)$ for the three-mode spin-boson model. The CPU simulation started with 50,000 walkers and the spawning generated $2.87 \times 10^7$ walkers in total. Accordingly, we allocated $2.8 \times 10^7$ walkers in the GPU simulation without spawning. The figure indicates that the accuracy is comparable between the CPU and GPU simulations. The computation time measured by the \texttt{@time} macro (excluding the compilation time) were $1860$ seconds on CPU and $80$ seconds on GPU.

\begin{figure}[htbp]
\centering
 \includegraphics[width=0.45\textwidth]{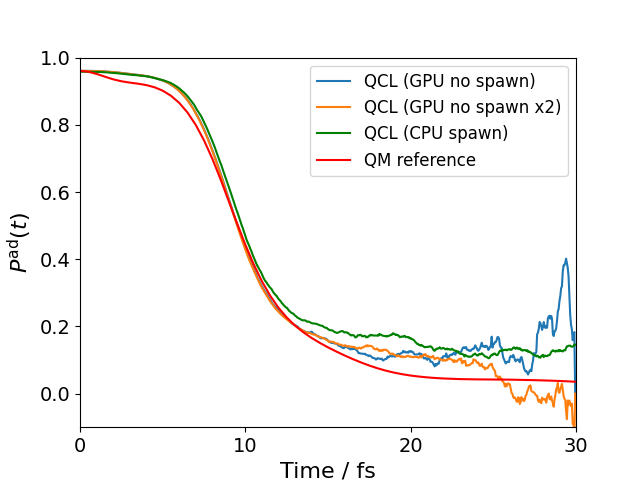}
 \caption{Adiabatic excited state population $P^{\text{ad}} \left(t\right)$ for three-mode model of pyrazine $S_2 \rightarrow S_1$ internal conversion, comparing results from GPU without trajectory spawning, CPU with trajectory spawning, and quantum reference calculation.}
 \label{fig:fig2}
\end{figure}

Figure 2 displays the corresponding results for pyrazine $S_2 \rightarrow S_1$ internal conversion. The CPU simulation started with 50,000 walkers and generated $1.40 \times 10^7$ walkers. The computation time was $540$ seconds. The GPU simulation with $1.4 \times 10^7$ walkers took $20$ seconds, but as seen in the figure, the deviation from the quantum reference was notable after 25 fs. We thus doubled the walkers to $2.8 \times 10^7$, which improved the accuracy as seen in the figure (see also Fig. \ref{fig:fig4} in Sec. \ref{sec:GPUscaling}). The computation time was $30$ seconds.

\subsection{ Scaling Performance on GPU}
\label{sec:GPUscaling} 
Here we examine the scaling properties of computation time to the number of walkers for the GPU calculation. Figure 3 corresponds to Fig. \ref{fig:fig1} but with varying numbers of walkers. The result with $N = 10^6$ is accurate enough up to 60 fs, but starts to deviate afterwards. The results up to 80 fs appear to have converged with $N = 10^7$.

The corresponding analysis for the pyrazine $S_2 \rightarrow S_1$ internal conversion is shown in Fig. \ref{fig:fig4}. Again, the simulation appears to have reasonably converged with $10^7$ walkers, but $10^6$ walkers would be sufficient to determine the decay rate.

Figure 5 displays the computation time {and the GPU memory usage} against the number of walkers, 
{both} indicating {linear scaling}. 
{A comparison between one- and three-mode spin-boson models also shows 
    linear scaling;
    the calculations with $N=10^8$ for one- and three-mode models
    took 115 and 268 seconds 
    and 5759 and 10943 MiB GPU memory, respectively. 
    Calculations with two GPUs also showed (nearly) linear scaling; 
the calculation of the three-mode spin-boson model with $N=10^8$ 
on two GPUs took 122 seconds and 5663 MiB.}

\begin{figure}[htbp]
\centering
 \includegraphics[width=0.45\textwidth]{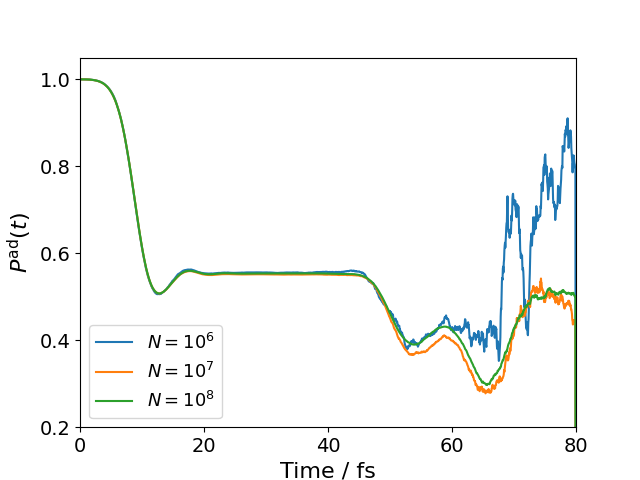}
 \caption{Adiabatic excited state population $P^{\text{ad}} \left(t\right)$ for three-mode spin-boson model from GPU simulation without trajectory spawning, comparing results with the number of walkers $N = 10^6 , 10^7$, and $10^8$.}
 \label{fig:fig3}
\end{figure}

\begin{figure}[htbp]
\centering
 \includegraphics[width=0.45\textwidth]{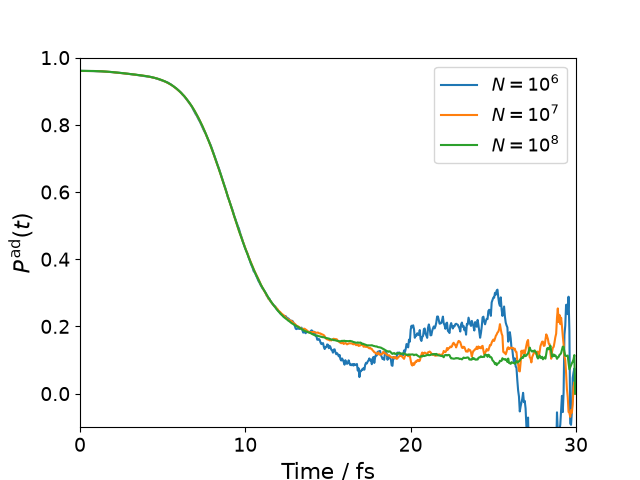}
 \caption{Adiabatic excited state population $P^{\text{ad}} \left(t\right)$ for three-mode model of pyrazine $S_2 \rightarrow S_1$ internal conversion, from GPU simulation without trajectory spawning, comparing results with the number of walkers $N = 10^6 , 10^7$, and $10^8$.}
 \label{fig:fig4}
\end{figure}

\begin{figure}[htbp]
\centering
 \includegraphics[width=0.40\textwidth]{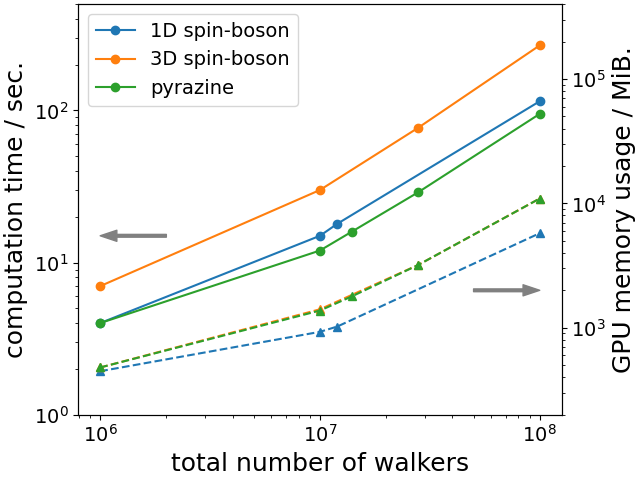}
 \caption{Log-log plot of computation time {(solid lines, left axis) and GPU memory usage (dashed lines, right axis)} 
     versus number of walkers 
 for the GPU simulation without trajectory spawning.}
 \label{fig:fig5}
\end{figure}

\section{ Concluding Remarks}
\label{sec:conclusion} We developed a GPU implementation of the momentum-jump-free mixed QCL MD simulation. To reduce GPU overheads related to thread divergence and dynamic memory allocation while keeping the code simplified, the trajectory spawning scheme previously used in CPU calculations was omitted. As a result, the GPU simulation achieved an order of magnitude speedup, although the exact factor would depend on the hardware specification. 
Furthermore, the computation time {and the GPU memory usage were} found to scale linearly with respect to the total number of simulated trajectories.
{They also scaled nearly linearly with the number of GPUs.
(However, further study with more GPUs on larger systems will be needed to confirm the scaling.)}

Because the nuclear motion in this framework is governed by the classical mechanics, 
the method can be seamlessly integrated into standard classical MD simulation for organic, inorganic, or biochemical systems, 
supplemented by additional computations of nonadiabatic couplings and multi-state forces. 
Due to its high parallel scalability, the present scheme is expected to benefit directly from 
ongoing developments in both hardware architectures (e.g., modern {many-}GPU {supercomputers} and unified memory) 
and software frameworks (such as QM/MM, fragment-based methods, multiple time-scale schemes). 
{Works on larger systems are currently in progress
and will be reported in future.}

\section*{Acknowledgments}
This work has been supported by JSPS KAKENHI Grant Number JP19K22173.

\section*{Data Availability}
The codes are available from github repository \\
(https://github.com/andok0208/QCLwomj.jl).

\section*{Declaration of generative AI and AI-assisted technologies in the manuscript preparation process}
During the preparation of this work, the author used Gemini for debugging the code.
The author reviewed and edited the output as needed and take full responsibility for the content of the published article.

\bibliographystyle{elsarticle-num}

\end{document}